\def\tr{{\rm tr}\,}

\def\b{\bibitem}

\documentstyle[aps,prl,floats,epsf]{revtex}
\draft
\begin{document}
% Macros for the various macro package names, etc.
\def\SNG{{\em Physical Review Style and Notation Guide}}
\def\LUG {{\em \LaTeX{} User's Guide \& Reference Manual}}
\def\btt#1{{\tt$\backslash$\string#1}}%
\def\REVTeX{REV\TeX}
\def\AmS{{\protect\the\textfont2
        A\kern-.1667em\lower.5ex\hbox{M}\kern-.125emS}}
\def\AmSLaTeX{\AmS-\LaTeX}
\def\BibTeX{\rm B{\sc ib}\TeX}
%\makeatletter
%\tighten
\twocolumn[\hsize\textwidth\columnwidth\hsize\csname@twocolumnfalse%
\endcsname
\title{Annealed disorder, rare regions, and local moments: A novel mechanism
       for metal-insulator transitions}
\author{D. Belitz,$^1$ T.R. Kirkpatrick,$^{2}$ and Thomas Vojta$^{1,3}$}
\address{$^1$Department of Physics and Materials Science Institute,
                               University of Oregon, Eugene, OR 97403}
\address{$^2$Institute for Physical Science and Technology, and Department
             of Physics, University of Maryland, College Park, MD 20742}
\address{$^3$Institut f{\"u}r Physik, TU Chemnitz, D-09107 Chemnitz, Germany}
\date{\today}
\maketitle
\begin{abstract}
It is shown that for noninteracting electron systems,
annealed magnetic disorder leads to a new mechanism, and a new universality
class, for a metal-insulator transition. The transition is driven by a 
vanishing of the thermodynamic density susceptibility rather than by
localization effects. The critical behavior in $d=2+\epsilon$ dimensions
is determined, and the underlying physics is discussed. It is further argued
that annealed magnetic disorder, in addition to underlying quenched disorder,
describes local magnetic moments in electronic systems.
%
%% 569 characters
%
\end{abstract}
]
%\narrowtext
Metal-insulator transitions (MITs) remain a fascinating and only incompletely
understood phenomenon\cite{R}. Conceptually, one distinguishes between
Anderson transitions in models of noninteracting electrons, and Mott-Hubbard 
transitions of clean, interacting electrons. At the former, the 
electronic charge diffusivity $D$ is driven
to zero by quenched, or frozen-in, disorder, while the thermodynamic
properties do not show critical behavior. At the latter,
the thermodynamic density susceptibility $\partial n/\partial\mu$ 
vanishes due to electron-electron interaction effects. In either case, the
conductivity $\sigma = (\partial n/\partial\mu)D$ vanishes at the MIT. In
many real systems both quenched disorder and interactions are present,
which makes a theoretical understanding of the resulting Anderson-Mott
transition very difficult. One particular complication is provided by
the presence of magnetic local moments (LMs) in such systems. There is much
experimental evidence for LMs\cite{LMs}, and their formation has been studied
theoretically\cite{Milovanovic}, but no existing theory can describe their 
interplay with the transport properties near the MIT\cite{R}.
Another complication is the possible presence of
{\em annealed} disorder, which is in thermal equilibrium with the rest of the 
system and hence involves disorder averaging of the partition function.
This is in contrast to quenched disorder that requires an averaging of
the free energy, which is usually done by means of the replica 
trick\cite{Grinstein}.

In this Letter we make two contributions to the MIT problem. (1) We show 
that annealed disorder
leads to a MIT that belongs to none of the previously studied classes.
It is driven by a vanishing $\partial n/\partial\mu$ and
thus resembles a Mott-Hubbard transition, even if no correlation effects
are explicitly considered. (2) We propose a mechanism by which
additional annealed disorder is generically self-generated in quenched 
disordered systems, and we argue that a type of LMs can be described in terms 
of it. We further develop a method for incorporating these `annealed LMs' 
into a transport theory.

Let us start by considering Wegner's nonlinear sigma-model 
(NL$\sigma$M)\cite{Wegner} for noninteracting electrons with nonmagnetic 
quenched disorder. The action reads
\begin{equation}
{\cal A} = \frac{-1}{2G}\int\! d{\bf x}\,\tr \left(\nabla
   Q({\bf x})\right)^2 + 2H\int\! d{\bf x}\,\tr \left(\Omega\,Q({\bf x})\right).
\label{eq:1}
\end{equation}
Here $Q({\bf x})$ is a matrix field that comprises two fermionic degrees of
freedom. Accordingly, $Q$ carries two Matsubara frequency indices $n$ and
$m$, and two replica indices $\alpha$ and $\beta$ to deal with the quenched
disorder. The matrix elements $Q_{nm}^{\alpha\beta}$ are spin-quaternion
valued to allow for particle-hole and spin degrees of freedom. It is
convenient to expand them in a basis $\tau_r\otimes s_i$ ($r,i=0,1,2,3$)
where $\tau_0=s_0$ is the $2\times 2$ unit matrix, and 
$\tau_{1,2,3} = - s_{1,2,3} = -i\sigma_{1,2,3}$, with $\sigma_j$ the Pauli
matrices. For simplicity, we will ignore the particle-particle or Cooper 
channel, which amounts to dropping $\tau_1$ and $\tau_2$ from the
spin-quaternion basis\cite{R}. $Q$ is subject to the constraints
$Q^2({\bf x}) \equiv 1$, and $\tr Q({\bf x}) \equiv 0$.
$\Omega_{nm}^{\alpha\beta} = \delta_{nm}\delta_{\alpha\beta}\Omega_n\,
(\tau_0\otimes s_0)$
is a frequency matrix with $\Omega_n = 2\pi Tn$ a bosonic Matsubara frequency
and $T$ the temperature.
$G$ is a measure of the disorder that is proportional 
to the bare resistivity, and the frequency coupling $H$ is proportional to
the bare density of states at the Fermi level. $\tr$ denotes a trace over
all discrete degrees of freedom that are not shown explicitly.

The properties of this model are well known\cite{Wegner,ELK,R}.
The bare action describes diffusive electrons, with $D=1/GH$ the diffusion 
coefficient.
Under renormalization $D$ decreases with increasing
disorder until a MIT is reached at a critical disorder value. The critical
behavior is known in an $\epsilon$-expansion about the lower 
critical dimension $d=2$. In the absence of the Cooper channel, the MIT
appears only at two-loop order at a critical disorder strength of
$O(\sqrt\epsilon)$. $H$, which determines the specific heat coefficient, 
the spin susceptibility, and $\partial n/\partial\mu$, is uncritical, 
which makes this MIT an Anderson transition.

Now we add magnetic annealed disorder to the model. Since our general
results are independent of its origin,
we first proceed without specifying it. Annealed disorder implies that
the $Q$ in the resulting terms all carry the same replica 
index\cite{Grinstein}; otherwise,
the functional form of the resulting additional
terms in the action can be taken from Ref.\ \onlinecite{ELK}. We obtain
two additional terms, viz.
\begin{mathletters}
\label{eqs:2}
\begin{eqnarray}
\Delta{\cal A}^{(1)}&=&\frac{TM_1}{8}\sum_{\alpha}\int d{\bf x}\sum_{j=1}^{3}
   \left[\tr\left(\left(\tau_3\otimes s_j\right)\,Q^{\alpha\alpha}({\bf x})
   \right)^2\right.
\nonumber\\
   &&\qquad\qquad\qquad - \left. \tr\left(Q^{\alpha\alpha}({\bf x})\right)^2
      \right]\quad,
\label{eq:2a}
\end{eqnarray}
and $\Delta{\cal A}^{(2)} = \Delta{\cal A}^{(2,s)} + \Delta{\cal A}^{(2,t)}$,
where 
\begin{eqnarray}
\Delta{\cal A}^{(2,j)}&=&\frac{TM_2^j}{8}
     \sum_{\alpha\neq\beta}
     \sum_{nm} \sum_{r=0,3} \int\! d{\bf x}\,\left[\tr (\tau_r\otimes s_i)\,
        Q_{nm}^{\alpha\beta}({\bf x})\right]
\nonumber\\
    &&\times \left[\tr (\tau_r^{\dagger}
               \otimes s_i)\,Q_{mn}^{\beta\alpha}({\bf x})\right]\quad,
\label{eq:2b}
\end{eqnarray}
\end{mathletters}%
with $j=s$ for $i=0$ (spin-singlet), and $j=t$ for $i=1,2,3$ (spin-triplet).
$\Delta{\cal A}^{(2)}$ arises from the need to absorb the scattering rate
due to the annealed disorder in $G$.
The coupling constants $M_1$, $M_2^s$, and $M_2^t$ are related to the strength 
of the magnetic annealed disorder. The factor of $T$ appears naturally
in front of any annealed disorder term, a crucial point that we will come back 
to later.

The action ${\cal A} + \Delta{\cal A}^{(1)}
+ \Delta{\cal A}^{(2)}$ can be analyzed by standard means. 
Note that the mass terms in Eqs.\ (\ref{eqs:2})
are proportional to temperature, making them quite different from conventional 
masses due to quenched disorder. In many respects, they are similar to 
electron-electron interaction terms in a Q-field theory 
formalism\cite{us_fermions}. We denote the
renormalized coupling constants that correspond to $G$, $H$, $M_1$, 
and $M_2^{s,t}$ by $g$, $h$, $m_1$, and $m_2^{s,t}$, 
and define $\delta_{1,2}^{s,t} = m_{1,2}^{s,t}/h$. The
renormalization group (RG) flow equations to one-loop order are
\begin{mathletters}
\label{eqs:3}
\begin{eqnarray}
\frac{dg}{dl}&=&-\epsilon g + g^2(\delta_2^s + 3\delta_2^t - 3\delta_1)\quad,
\label{eq:3a}
\\
\frac{dh}{dl}&=&-hg(\delta_2^s + 3\delta_2^t - 3\delta_1)\quad,
\label{eq:3b}
\\
\frac{d\delta_1}{dl}&=&-g\left[-4\delta_1^2 + \delta_1(\delta_2^s + 3\delta_2^t)
   + (\delta_2^s - \delta_2^t)^2\right]\quad,
\label{eq:3c}
\\
\frac{d\delta_2^s}{dl}&=&g\left[3\delta_1^2 
   + 3\delta_1(\delta_2^s - 2\delta_2^t)
   - 3\delta_2^t(\delta_2^s - \delta_2^t)\right]\quad,
\label{eq:3d}
\\
\frac{d\delta_2^t}{dl}&=&g\left[3\delta_1^2 - \delta_1(2\delta_2^s + \delta_2^t)
   -(\delta_2^s)^2 \right.
\nonumber\\
&&\qquad\qquad\qquad-\left. 2(\delta_2^t)^2 + 3\delta_2^s\delta_2^t\right]\quad,
\label{eq:3e}
\end{eqnarray}
\end{mathletters}%
where $l=\ln b$ with $b$ the RG length scale factor. Besides unstable fixed
points (FPs), there is a line of critical fixed points (FPs) 
$(g^*,h^*,\delta_1^*,\delta_2^{s*},
\delta_2^{t*}) = (\epsilon/4\delta_2^*,0,0,\delta_2^*,\delta_2^*)$ that
correspond to an MIT (all of these FPs belong to the same universality
class). Linearization about any of these FPs yields one relevant eigenvalue
$\lambda_g = \epsilon + O(\epsilon^2)$ that determines the correlation 
length exponent $\nu = 1/\lambda_g$, one marginal eigenvalue that 
corresponds to moving along the line of FPs, and two irrelevant 
eigenvalues equal to $-\epsilon + O(\epsilon^2)$.
The anomalous dimension of $h$ is $\kappa = -\epsilon + O(\epsilon^2)$.
In addition, the critical behavior of the single-particle density of states
(DOS), $N$, at the Fermi level can be obtained from the 
wavefunction renormalization.
Choosing the critical exponent of the DOS, $\beta$, the correlation length
exponent $\nu$, and the dynamical critical exponent $z = d + \kappa$ as
independent exponents, we find
\begin{equation}
\nu = 1/\epsilon + O(1)\ ,\ \beta = \epsilon + O(\epsilon^2)\ ,\ 
   z = 2 + O(\epsilon^2)\ .
\label{eq:4}
\end{equation}
For the conductivity exponent we find $s=\nu\epsilon = 1 + O(\epsilon)$,
and $\partial n/\partial\mu$, the spin susceptibility $\chi_s$, and the
specific heat coefficient $\gamma = C_V/T$, which we collectively denote
by $\chi$, all vanish with a critical
exponent determined by $\kappa$. The diffusion coefficient, on the other hand,
has no anomalous dimension and thus
is uncritical to one-loop order, as can be seen from Eqs.\ (\ref{eq:3a},
\ref{eq:3b}). 
With $t$ the dimensionless distance from the critical point at $T=0$,
and $E$ the energy, we can summarize the critical behavior of these
quantities by the homogeneity laws
\begin{mathletters}
\label{eqs:5}
\begin{eqnarray}
\chi(t,T)&=&b^{\kappa} \chi(tb^{1/\nu},Tb^z)\quad,
\label{eq:5a}\\
N(t,T,E)&=&b^{-\beta/\nu} N(tb^{1/\nu},Tb^z,Eb^z)\quad,
\label{eq:5b}\\
\sigma(t,T)&=&b^{-s/\nu} \sigma(tb^{1/\nu},Tb^z)\quad,
\label{eq:5c}\\
D(t,T)&=&b^{-(s/\nu+\kappa)} D(tb^{1/\nu},Tb^z)\quad.
\label{eq:5d}
\end{eqnarray}
\end{mathletters}%
We conclude that the MIT is driven by the vanishing of
$\partial n/\partial\mu$, and therefore is qualitatively different from
the localization transition that is found in the absence of annealed
disorder. Indeed, putting $M_1 = M_2^s = M_2^t = 0$ we find that all
thermodynamic anomalies disappear, as does the one-loop correction to
$g$. At two-loop order, one finds instead a MIT of Anderson type\cite{R}.

We now turn to a specific realization, via local magnetic moments,
of the annealed disorder that leads to the
striking effects discussed above. To explain the salient points,
it is easiest to initially consider a simpler field theory than the
$Q$-matrix theory studied above, and adapt a classical line of reasoning
from Ref.\ \onlinecite{Dotsenko} to quantum field theories. Accordingly, 
we consider a scalar quantum field $\phi({\bf x},\tau)$ and an action 
\begin{equation}
S[\phi] = \int dx\,\left(\phi\partial_{\tau}\phi
          - {\cal H}[\phi,\nabla\phi]\right)\quad,
\label{eq:6}
\end{equation}
Here $x=({\bf x},\tau)$ comprises position ${\bf x}$ and imaginary time 
$\tau$, $\int dx \equiv \int d{\bf x}\int d\tau$, 
${\cal H}$ is a Hamiltonian density, and
we use units such that $\hbar=k_{\rm B}=1$. We will assume that $S$
describes a phase transition from a disordered to an ordered phase,
and will use a magnetic language, referring to $\langle\phi\rangle$ as
`magnetization'. Suppose that ${\cal H}$ contains
quenched disorder of random-mass type, and that we are in the nonmagnetic
phase, $\langle\phi\rangle=0$. The key idea is to {\em not} integrate out
the quenched disorder as a first step, as one does in a conventional
treatment\cite{Grinstein}, but rather to work with a particular disorder
realization. Due to the quenched disorder there will
be regions in space that energetically favor local order, 
$\langle\phi\rangle\neq 0$, even though there is no global order. Deep
inside the disordered phase these regions will be rare, but in an
arbitrarily large system we will find arbitrarily large rare regions 
with a finite probability. The action $S$ will then have static saddle-point
solutions $\Phi({\bf x})$ that have a nonvanishing value of the magnetization 
only in
the rare regions. Let there be $N$ such rare regions and associated
local blobs of magnetization or LMs. Then we can
actually construct $2^N$ such saddle points, which differ only by the
way the sign of the magnetization is distributed among the LMs.
Since the LMs are far apart, the energy differences between these
$2^N$ saddle points will be small. In expanding about
the saddle points, we therefore have no reason to prefer one of them
over any of the others. Furthermore, since the LMs are self-generated
by the system, albeit in response to the quenched potential, we assume
that they are in thermal equilibrium with all other degrees of freedom
as well as with each other. To calculate the partition function $Z$ it is 
therefore necessary to take into account fluctuations in the vicinity of 
each of the $2^N$ saddle points\cite{barrier_footnote}:
\begin{equation}
Z \approx \sum_{a=1}^{2^N} \int_{<} D[\varphi]\ 
            \exp \left(-S[\Phi^{(a)} + \varphi]\right)\quad.
\label{eq:7}
\end{equation}
Here $\int_{<} D[\varphi]$ denotes an integration over small fluctuations
$\varphi$
in the vicinity of each of the saddle points. Notice that this restriction
to small fluctuations is necessary in order to avoid double counting.
Conversely, if we could perform the integral over the fluctuations exactly,
then it would be sufficient to expand about one of the saddle points. In
practice, however, one is restricted to a perturbative evaluation of the
functional integral, and Eq.\ (\ref{eq:7}) is a good 
approximation\cite{nonperturbative_footnote}.

We now consider the thermodynamic limit. Then the discrete set of $2^N$
saddle points turns into a saddle-point manifold ${\cal M}(\Phi)$ that
needs to be integrated over. Splitting off the saddle-point part of the action, 
$S[\phi] = S[\Phi] + \Delta S[\Phi,\varphi]$, we have
\begin{mathletters}
\label{eqs:8}
\begin{equation}
Z = \int D[\Phi]\ P[\Phi]\int D[\varphi]\ 
                    \exp\left(-\Delta S[\Phi,\varphi]\right)\quad,
\label{eq:8a}
\end{equation}
with the probability distribution $P$ given by
\begin{equation}
P[\Phi] = {\cal S}(\Phi)\ \exp\left(-\frac{1}{T}\int d{\bf x}\ 
 {\cal H}[\Phi,\nabla\Phi]\right)\quad.
\label{eq:8b}
\end{equation}
\end{mathletters}%
Here ${\cal S}$ denotes the support of the saddle-point manifold ${\cal M}$.
Notice the factor of $1/T$ in the exponent, which results from the static
nature of the saddle points\cite{sign_footnote}.

In general it is not possible to determine $P[\Phi]$ explicitly.
However, if we perform the $\Phi$ integration by 
means of a cumulant expansion, the 
most relevant term in the effective action will be the one that
results from the term quadratic in ${\cal H}[\Phi]$ and the linear coupling
between $\Phi$ and $\varphi^2$ in $\Delta S$. To obtain the most relevant
term in the effective theory for the fluctuations $\varphi$, we thus can
write, with $w>0$ a number\cite{sign_footnote},
\begin{eqnarray}
Z&\approx&\int D[\varphi]\ e^{-S[\varphi]}\int D[\Phi]\ 
     \exp\left(\frac{-1}{wT}\int d{\bf x}\, \Phi^2({\bf x})\right)
\nonumber\\
&&\times \exp\left({\int dx\ \Phi({\bf x})\,
     \varphi^2(x)}\right)\quad.
\label{eq:9}
\end{eqnarray}

Equation (\ref{eq:9}) is the partition function one would obtain by
expanding perturbatively about just one of the saddle points, 
with static, annealed
disorder appearing in addition to the quenched disorder still contained
in $S[\varphi]$. The annealed disorder is governed by a Gaussian
distribution whose variance is proportional to $T$. This property 
reflects the fact that the annealed disorder, as classical degrees 
of freedom in equilibrium with the rest of the system, must come with a
Boltzmann weight, and it is the reason for the factors of $T$ in
Eqs.\ (\ref{eqs:2}).

Let us now explain how these arguments can be applied to the $Q$-field
theory of interacting electrons to arrive at the action, Eqs.\ (\ref{eq:1},
\ref{eqs:2}). The magnetization is proportional to the expectation value 
$\langle\tr (\tau_3\otimes s_i)\,Q({\bf x})
\rangle$ \cite{us_fermions}, and in the presence of quenched disorder that
favors the formation of magnetic LMs, the exact fermionic theory
that underlies the NL$\sigma$M\cite{us_fermions} allows for
saddle-point solutions where these components of $Q$ are locally nonzero
and play the role of the field $\Phi$ above. This is
in addition to a globally nonzero $\langle\tr (\tau_0\otimes s_0)\,Q({\bf x})
\rangle$ which reflects a nonvanishing DOS.
By following the above reasoning for a scalar field,
and going through the derivation of the sigma-model again, one obtains
Eqs.\ (\ref{eq:1},\ref{eqs:2}). 

We conclude with several remarks. First, we emphasize that we have studied
a simplified model, neglecting both the Cooper channel and the
electron-electron interaction. The latter point requires some clarification.
In order to generate the annealed disorder from LMs, some interaction is 
necessary, (1) for local magnetic order to develop, and (2) in order for 
our canonical averaging over the saddle points to 
make physical sense. A truly noninteracting system would not sample all 
of these field configurations. Put differently,
interactions make the energy barriers between the saddle points, which
are infinite in a noninteracting system, finite and thus
allow for an equilibration of the saddle-point degrees of 
freedom\cite{barrier_footnote,nonperturbative_footnote}. We have
simplified our model by assuming points (1) and (2) above to be the {\em only}
effect of the interactions. Of course, if the annealed disorder were due to
some other mechanism, then our results would also apply to strictly
noninteracting electrons. Clearly, one can study generalizations
of our model. In addition to adding an explicit interaction term, one can
restore the Cooper channel, which will make the FP we found compete with
the ordinary localization FP that also occurs at one-loop order. In systems
with time reversal symmetry, one then
expects the MIT studied here to get preempted by a localization transition
if the bare dimensionless mass $M_2/H$ is smaller than a number of $O(1)$.
It would also be interesting to consider the present model to 2-loop order
to see whether the diffusion coefficient will still not be renormalized
(apart from the `diffuson' localization contributions that will appear at
that order), 
and whether the line of FPs gives way to a more conventional FP structure.
These questions will be considered in the future.

Second, we point out that the strong effects of annealed
disorder we found are characteristic of quantum statistical mechanics.
In a classical scalar field theory, the leading term in the 
action generated upon 
integrating out annealed disorder is of the form (see Eq.\ (\ref{eq:9}))
$-\int d{\bf x}\,\varphi^4({\bf x})$. It thus has the same form as the
ordinary $\varphi^4$-term and is in general not very interesting (although it
can lead, e.g., to a first order phase transition). 
In a quantum system, on the other hand, integrating
out the annealed disorder yields
$-\int d{\bf x}\int d\tau\,d\tau'\,\varphi^2({\bf x},\tau)\,
\varphi^2({\bf x},\tau')$, which has a different time structure than the
usual $\varphi^4$ term. It is the extra time integral that makes the
annealed disorder term more relevant than in the classical case.

Third, we come back to the fact that the variance of the Gaussian distribution
for the annealed disorder is linear in $T$. If one used a Gaussian distribution
with a temperature independent width, one would encounter factors of $1/T$
in perturbation theory that force one to scale the annealed disorder strength
with $T$ to obtain a meaningful theory. Annealed disorder with an unbounded
distribution and a finite variance at $T=0$ is unphysical, since it allows
the system to lower its energy arbitrarily far by digging itself a deeper and
deeper trough. The necessity of the factor of $T$ was realized in 
Ref.\ \onlinecite{rr_magnets}, but its
origin was not recognized\cite{1/T_footnote}.

Finally, let us explain why annealed disorder leads to a critical
$\partial n/\partial\mu$, while quenched disorder without electron-electron
interactions does not. To see this, we realize that annealed disorder
essentially means potential troughs that are somewhat flexible, i.e. they
adjust in response to the electrons. Let the sytem be in
equilibrium at some value of the chemical potential $\mu$, and
change $\mu$ slightly. Then the flexible potential will adjust, and
as a result fewer electrons will have to flow out of or into the grand
canonical reservoir than would be the case in the absence of annealed
disorder. This explains why there is a correction to $\partial n/\partial\mu$
in perturbation theory. Furthermore, the diffusive dynamics of the electrons
lead to this correction being a frequency-momentum integral over diffusion
propagators, which is logarithmically singular in $2$-$d$. In $d=2+\epsilon$
this leads to a critical $\partial n/\partial\mu$, as it happens with other
quantities that are singular in perturbation theory in $2$-$d$. This is the
only known mechanism for a critical $\partial n/\partial\mu$ at a MIT in
low-dimensional systems\cite{high-d}. The recent observation
of a critical $\partial n/\partial\mu$ at a $2$-$d$ MIT\cite{Jiang} 
is therefore very interesting in this context, even though our current 
theory does not describe a MIT in $d=2$.

This work was initiated at the Aspen Center for Physics, and
supported by the NSF under grant Nos. DMR-98-70597 and
DMR--96--32978, and by the DFG under grant No. SFB 393/C2.


\vskip -4mm
\begin{references}
\b{R} For a review, see, e.g., D. Belitz and T.R. Kirkpatrick, Rev. Mod.
 Phys. {\bf 66}, 261 (1994). 
\b{LMs} M.A. Paalanen, S. Sachdev, R.N. Bhatt, and A.E. Ruckenstein, Phys.
 Rev. Lett. {\bf 57}, 2061 (1986); M.A. Paalanen, J.E. Graebner, R.N. Bhatt,
 and S. Sachdev, Phys. Rev. Lett. {\bf 61}, 597 (1988); Y. Ootuka and
 N. Matsunaga, J. Phys. Soc. Japan {\bf 59}, 1801 (1990).
\b{Milovanovic} M. Milovanovic, S. Sachdev, and R.N. Bhatt, Phys. Rev. Lett.
 {\bf 63}, 82 (1989), and references therein.
\b{Grinstein} See, e.g., G. Grinstein, in {\it Fundamental Problems in
 Statistical Mechanics VI}, E.G.D. Cohen (ed.), North Holland (Amsterdam,
 1985).
\b{Wegner} F. Wegner, Z. Phys. B {\bf 35}, 207 (1979). We use the fermionic
 formulation of the model given in Ref.\ \onlinecite{ELK}. See also
 Ref.\ \onlinecite{us_fermions}.
\b{ELK} K.B. Efetov, A.I. Larkin, and D.E. Khmelnitskii, Zh. Eksp. Teor. Fiz.
 {\bf 79}, 1120 (1980) [Sov. Phys. JETP {\bf 52}, 568 (1980)].
\b{us_fermions} D. Belitz and T.R. Kirkpatrick, Phys. Rev. B {\bf 56}, 6513 
 (1997); D. Belitz, T.R. Kirkpatrick, and F. Evers, Phys. Rev. B {\bf 58},
 9710 (1998).
\b{Dotsenko} Viktor Dotsenko, A.B. Harris, D. Sherrington, and R.B. Stinchcombe,
 J. Phys. A {\bf 28}, 3093 (1995).
\b{barrier_footnote} Elsewhere we have argued that for noninteracting LMs, 
 the typical barriers between different saddle point states diverge 
 in the bulk limit\cite{rr_magnets}. We expect that one effect of interactions 
 between the LMs is to make the barriers finite, so that the
 averaging discussed here applies.
\b{rr_magnets} R. Narayanan, T. Vojta, D. Belitz, and T.R. Kirkpatrick, Phys,
 Rev. B {\bf 60}, 10150 (1999).
\b{nonperturbative_footnote} The basic assumption underlying these arguments
 is that typical pairs of saddle-point configurations are separated by finite
 barriers\cite{barrier_footnote}, but not perturbatively accessible from one 
 another. This amounts to an assumption about a separation of time scales
 that we expect to be true in some, but not all, systems.
\b{sign_footnote} If one uses a simple saddle-point theory of noninteracting
 LMs for $\Phi$ in the Hamiltonian density in Eq.\ (\ref{eq:8b}),
 then the coefficient of $1/T$ is found to be positive, indicating that the 
 LM state is a thermodynamic equilibrium state. We expect a more
 realistic treatment, including LM interactions, to weaken the 
 effects of these LMs and make the state thermodynamically unstable, 
 changing the sign of the $1/T$ term to the one used in 
 Eqs.\ (\ref{eq:8b},\ref{eq:9}), which physically makes these 
 LMs {\em dynamic} rare fluctuations.
\b{1/T_footnote} Developing the theory without the explicit 
 factor of $T$ in Eqs.\ (\ref{eqs:2}) leads to
 Eqs.\ (\ref{eqs:3}) with an extra factor of $1/T$ on the righ-hand side. 
 The only physical fixed point then requires the masses to scale with
 $T$. The factor of $T$ in Eqs.\ (\ref{eqs:2}) is thus unavoidable on
 technical grounds, even if its presence on physical grounds is not realized.
\b{high-d} In high dimensions an order parameter theory of the Anderson-Mott
 transition yields a critical $\partial n/\partial\mu$, T.R. Kirkpatrick
 and D. Belitz, Phys. Rev. Lett. {\bf 73}, 862 (1994), while in  
 $d=2+\epsilon$, $\partial n/\partial\mu$ is uncritical\cite{R}.
\b{Jiang} S.C. Dultz and H.W. Jiang, cond-mat/9909314.
\end{references}
\end{document}